\documentstyle[11pt,newpasp,twoside,epsf]{article}
\begin{document}
\title{Dust and Molecular Emission from the S235 Star Forming Region}
 \author{Jeong-Eun Lee, Chadwick H. Young, Yancy L. Shirley, Kaisa E. Mueller, \& 
	Neal J. Evans II }
\affil{University of Texas at Austin, Austin, Texas 78712-1083}

\begin{abstract}
We present the results of a multi-faceted study of the high-mass star-forming 
region S235. With the CalTech Submillimeter Observatory (CSO), we have observed 
and/or mapped over fifty molecular transitions and 350 $\micron$ dust emission 
in an attempt to learn more about the chemistry, kinematics, and mass 
distribution within this rich region. 
We have found two dust cores that are not resolved in our molecular line observations
and suggest  that one of these is a less evolved star-forming core. 
\end{abstract}

\section{Introduction}
The S235 star-forming region includes small optical nebulosities (S235A, S235B,
and S235C), and it is located about 10$\arcmin$ south of S235, which is a well 
known HII region.
This star-forming region has been identified by $\rm H_2O$ maser (Lo et al. 
1975, Genzel \& Downes 1977), infrared emission (Krassner et al.  1997, 
Evans et al. 1981, Felli et al.1997), radio continuum (Krassner et al. 1979,
Felli et al. 1997), and molecular line spectra (Evans et al. 1981, Makoto et al.
1986, Snell et al. 1990, Felli et al. 1997). According to previous results,
the region has several components with somewhat different spatial distributions.
We present the initial findings of a major observing campaign aimed at understanding
the complexity of S235 with greater lucidity.
        
\section{Dust Continuum Emission and Molecular Spectra}

We observed the dust emission from S235 at 350 $\micron$ with the Submillimeter High 
Angular Resolution Camera (SHARC) at the CSO during December 1997 and show the 
resulting map, with a resolution of about 11\arcsec, in Figure 1.
There are two dust cores; the northern one is close to and, presumably, related to the compact 
HII region, S235A, and the infrared source, IRS3.
Krassner et al. (1979) found two 2.2 $\micron$ point source in S235A, 
and Felli et al. (1997) observed molecular hydrogen emission distributed 
around the HII region. This $\rm H_2$ emission is along the 
southeast edge of our northern dust core.
Several infrared point sources observed at 2.2$\micron$ (Felli et al. 1997) are 
located between the two dust cores. 
On the other hand, the southern dust core is not associated with any 
optical nebula, HII region, or NIR point source to heat it. 
S235B and IRS4, the nearest signposts of activity,  are about 0.3 pc away from the peak of this core.
Therefore, we suggest that this is a less evolved star-forming core.

We observed over 50 transitions of 23 molecules ($\rm CS$, $\rm C^{34}S$, $\rm
^{13}CS$, $\rm C^{17}O$, $\rm C^{18}O$, $\rm HCN$, $\rm H^{13}CN$, $\rm HC^{15}N$,
$\rm DCN$, $\rm HNC$, $\rm HN^{13}C$, $\rm DNC$, $\rm HCO^+$, $\rm H^{13}CO^+$, $\rm HC^{18}O^+$, $\rm DCO^+$,
$\rm HCS^+$, $\rm SO$, $\rm CH_3CCH$, $\rm CH_3CN$, $\rm CH_3OH$, $\rm H_2CO$, 
and $\rm SO_2$) at the CSO from 1998 to 2000. The typical line width of 
these molecular spectra is about 3kms$^{-1}$ at the peak for the northern molecular core.  
The CS $J=5\rightarrow4$ integrated intensity map (Figure 1) clearly resolves
two molecular cores, of which the southern core is less pronounced.  
The two dust cores, however, are not resolved due to poorer angular resolution (25\arcsec), 
but we do see a slight extension of the northern molecular core in the direction of the southern dust core.
Further molecular line observations with better angular resolution 
are needed to study gas components associated with the two dust components.

\begin{figure}
\plotfiddle{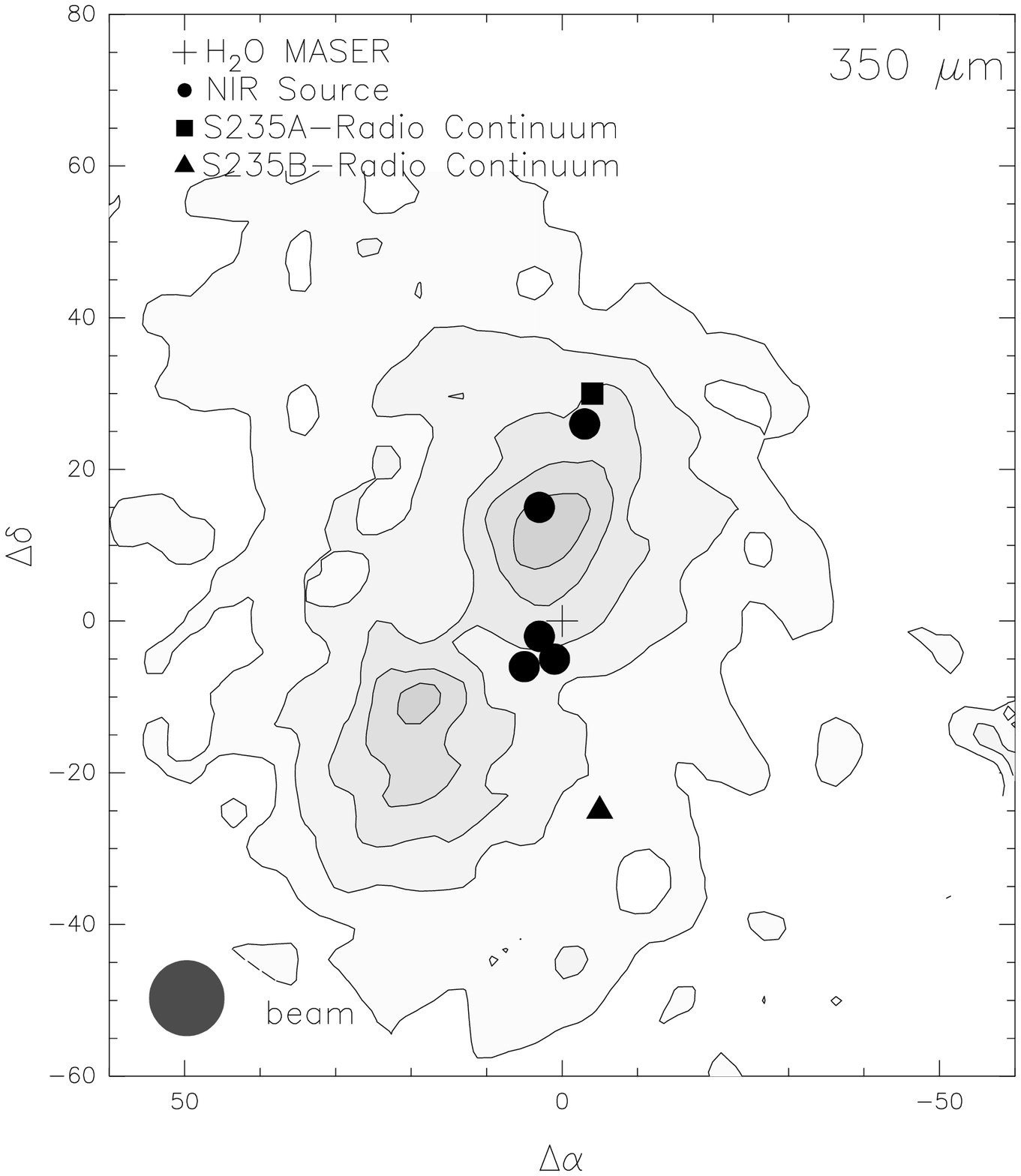}{150pt}{0}{30}{30}{0}{-15}
\plotfiddle{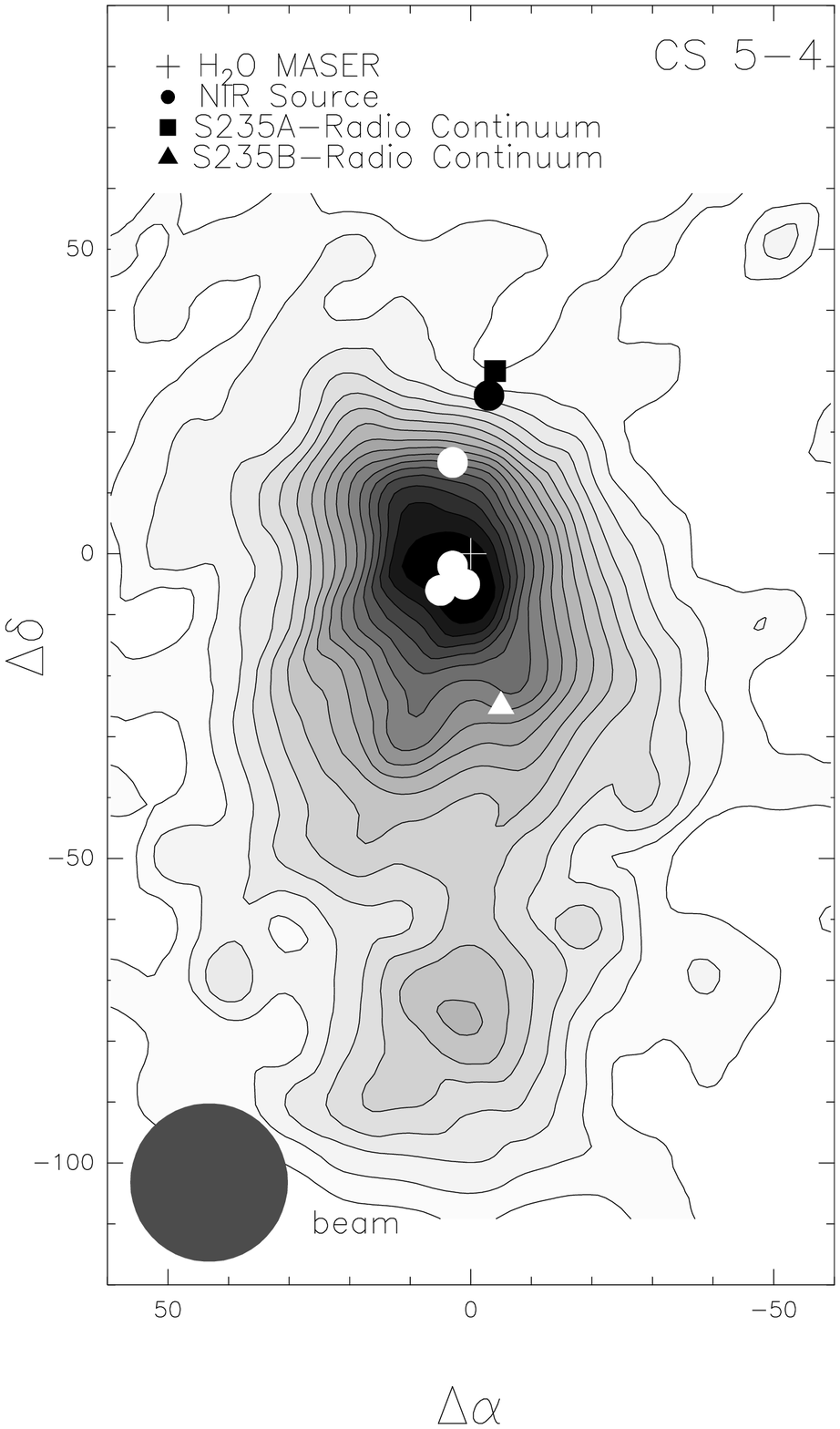}{0pt}{0}{30}{30}{-150}{-15}
\caption{In the left figure, we show the CS$5\rightarrow4$ map of S235  and, on the right, 
a map of the dust continuum emission. The central position for each is $\alpha_{1950}=05h37m31.8s$
and $\delta_{1950}=35\deg40\arcmin18\arcsec$.  Note that the two peaks in the dust map are not resolved by our
molecular line observations.}
\end{figure}

\end{document}